\documentclass[%
 aip,
 amsmath,amssymb,
 reprint,%
]{revtex4-2}

\usepackage{graphicx}
\usepackage{dcolumn}
\usepackage{bm}
\usepackage[utf8]{inputenc}
\usepackage[T1]{fontenc}
\usepackage{mathptmx}
\usepackage{etoolbox}
\usepackage{xcolor}

\makeatletter
\def\@email#1#2{%
 \endgroup
 \patchcmd{\titleblock@produce}
  {\frontmatter@RRAPformat}
  {\frontmatter@RRAPformat{\produce@RRAP{*#1\href{mailto:#2}{#2}}}\frontmatter@RRAPformat}
  {}{}
}%
\makeatother
\begin{document}

\preprint{AIP/123-QED}

\title{Chemically symmetric and asymmetric self-driven rigid dumbbells in 2D polymer gel}
\author{Praveen Kumar}
 \affiliation{Department of Chemistry, Indian Institute of Technology Bombay, Mumbai, Maharashtra -  400076, India}
\author{Ligesh Theeyancheri}
\affiliation{Department of Chemistry, Indian Institute of Technology Bombay, Mumbai, Maharashtra -  400076, India}
\author{Rajarshi Chakrabarti$^\ast$}
\email{rajarshi@chem.iitb.ac.in}
\affiliation{Department of Chemistry, Indian Institute of Technology Bombay, Mumbai, Maharashtra -  400076, India}

\begin{abstract}
\noindent We employ computer simulations to unveil the translational and rotational dynamics of the self-driven chemically symmetric and asymmetric rigid dumbbells in two-dimensional polymer gel. Our results show that activity or the self-propulsion always enhances the dynamics of the dumbbells. Making the self-propelled dumbbell chemically asymmetric leads to further enhancement in dynamics. Additionally, the direction of self-propulsion is a key factor for the chemically asymmetric dumbbells, where self-propulsion towards the non-sticky half of the dumbbell results in faster translational and rotational dynamics compare to the case with the self-propulsion towards the sticky half of the dumbbell. Our analyses show that both the symmetric and asymmetric passive rigid dumbbells get trapped inside the mesh of the polymer gel, but the chemical asymmetry always facilitates mesh to mesh motion of the dumbbell and it is even more pronounced when the dumbbell is self-propelled. This results multiple peaks in the van Hove function with increasing self-propulsion. In a nutshell, we believe that our in silico study can guide the researchers design efficient artificial microswimmers possessing potential applications in site-specific delivery.
\end{abstract}

\maketitle

\section{Introduction}\label{Intro}

\noindent Biological processes such as cell motility, cell division, and cellular transport are governed by non-equilibrium events like force generation by ATP hydrolysis and persistent motion of the individual constituents inside the complex and dense physiological environments. For example, motility of spermatozoa by ciliary beating through cervical mucus \citep{riedel2005self,fauci2006biofluidmechanics}, spindle oscillations during asymmetric cell division \citep{pecreaux2006spindle}, microproteins in cells, and microbial pathogens in stomach mucus \citep{celli2009helicobacter}, all indicate that the living matter undergoes different non-equilibrium events. Inspired by these biological swimmers, researchers have come up with artificial microswimmers mimicking the role of their biological counterparts. For instance, half-coated Janus colloids \citep{howse2007self,gomez2016dynamics}, catalytic nanomotors \citep{wang2015fabrication}, chiral particles \citep{ghosh2009controlled}, and vesicles \citep{joseph2017chemotactic} are able to self-propel by different propulsion mechanisms like diffusiophoresis \citep{golestanian2005propulsion,palacci2010sedimentation}, bubble propulsion \citep{gibbs2009autonomously}, thermophoresis \citep{jiang2010active}, self-electrophoresis \citep{wang2006bipolar}\textit{etc}. Macromolecular crowding plays a vital role in deciding the motion of active agents through the physiological environments like living cells where the biomolecules are transported through the networks formed by biofilaments~\citep{lombardo2019myosin} and molecules diffuse through the chromatin networks inside the eukaryotic nuclei~\citep{fritsch2010anomalous}. The biomolecular transport depends on the effective diffusion of biological swimmers of different chemical or mechanical properties, and while moving through the confined dense space, they encounter collisions and experience range of interactions from the media, which greatly influence their dynamics. Thus motility of microswimmers in crowded environments is controlled by both the self-propulsion force, the random force exerted by the media, and the interactions with the media. \\

\noindent In the recent past, a number of experimental and theoretical studies have been carried out to investigate the dynamics of the self-propelled species in crowded and complex environments \citep{thapa2019transient, sakha2010three, wang2016target, joo2020anomalous, lowen2020inertial, theeyancheri2020translational, chaki2020escape, gera2021solution, caprini2021correlated, goswami2022motion}. Recent experimental and theoretical studies reported enhanced rotation of a self-propelled Janus particle in polymer solution \citep{gomez2016dynamics,theeyancheri2020translational, singh2022interaction}. Most of the previous attempts focus on the effect of crowding on dynamics by changing the nature and density of the crowders~\citep{joo2020anomalous,sakha2010three,wang2016target, goodrich2018enhanced, kumar2019transport, chen2022passive}. However, much less is explored on the effect of shape and chemical behavior of the microswimmers on its dynamics in complex media \citep{khalilian2016obstruction}. Anisotropy in shape and interaction leads to more complicated dynamical behaviors \citep{theeyancheri2020translational, khalilian2016obstruction}. The presence of crowders amplifies the difference between the dynamical behavior of an asymmetric and symmetric species. Computational studies of infinitely thin needles in a two-dimensional array of point obstacles have shown that the transport is increasingly faster at higher densities, while the thicker needles suppress the enhancement in dynamics \citep{tucker2010observation, hofling2008enhanced}. A more fundamental question would be to ask, what happens if the microswimmer possesses chemical asymmetry and how the dynamics differ from the symmetric analog inside a dense medium. Researchers have synthesized chemically asymmetric Janus rods and chiral particles for various applications, and these artificial microswimmers or the self-powered nano/microdevices are considered as next-generation drug delivery systems\citep{patra2013intelligent, sundararajan2008catalytic}. So a deeper understanding of the macromolecular crowding effects induced by the environment and symmetric-asymmetric dependence on their dynamics is absolutely necessary to build novel strategies for artificial microswimmers with better performance, which has broad importance and potential applications in diverse fields including cellular biophysics, nanomedicine, drug delivery, and so on.     \\

\noindent In the present work, we numerically explore the dynamics of the chemically symmetric and asymmetric self-propelled rigid dumbbells inside a polymer gel. The dumbbells are made of two circular particles connected by a spring, where the self-propulsion always acts along the spring (Fig.~\ref{fig:2D_Gel}). The gel is constructed on a 2D graphene lattice, where each site is occupied by a particle and these particles are connected by springs (Fig.~\ref{fig:2D_Gel}A). We analyze both the translational and rotational dynamics of the chemically symmetric and asymmetric self-propelled rigid dumbbells in detail by tracking the center of mass (COM) and the coordinates of two halves of the rigid dumbbells respectively, as it is usually done in single particle tracking experiments \citep{anthony2008single, anthony2015tracking, rose2020particle}. Our results show that the self-propulsion always enhances the dynamics and the chemical asymmetry plays a pivotal role in determining the rotational dynamics of the self-propelled rigid dumbbells. For the chemically symmetric dumbbells, both the halves are attractive (sticky) to the gel particles. On the other hand, for chemically asymmetric dumbbells, one half is sticky and the other half is repulsive (non-sticky) to the gel particles. We found that the chemical asymmetry of the rigid dumbbell results additional enhancement in rotational motion compared to its symmetric cousin. Adding to this, the direction of self-propulsion has been found to be a crucial factor in case of asymmetric rigid dumbbell, as the self-propulsion towards the non-sticky face leads to enhanced translational and rotational dynamics in comparison to the self-propulsion towards the sticky face. Apart from this, we observe that the passive rigid dumbbell is trapped inside the mesh confinement of the gel irrespective of whether chemically symmetric or asymmetric. Contrary to that, the asymmetric dumbbell moves from one mesh to the adjacent mesh smoothly compared to the symmetric dumbbell. Interestingly, the activity or the self-propulsion helps the dumbbell move from one mesh to another and explore a larger space inside the polymer gel, which results in the multiple numbers of oscillations in the van Hove function with increasing activity. This mesh to mesh movement of the self-driven dumbbell is more pronounced for higher activities independent of the direction of self-propulsion and the interaction with the polymer gel. This implies that activity dominates over interaction strength at larger values of self-propulsion. We believe that the present investigation will help the researchers design more efficient self-powered microswimmers by incorporating chemical asymmetry and controlling the direction of self-propulsion. This would have wide applications in the field of nanomedicine for the site-specific delivery of drugs.\\

\section{Model and simulation details}\label{Model}

\begin{figure*}
    \centering
    \includegraphics[width=0.9\linewidth]{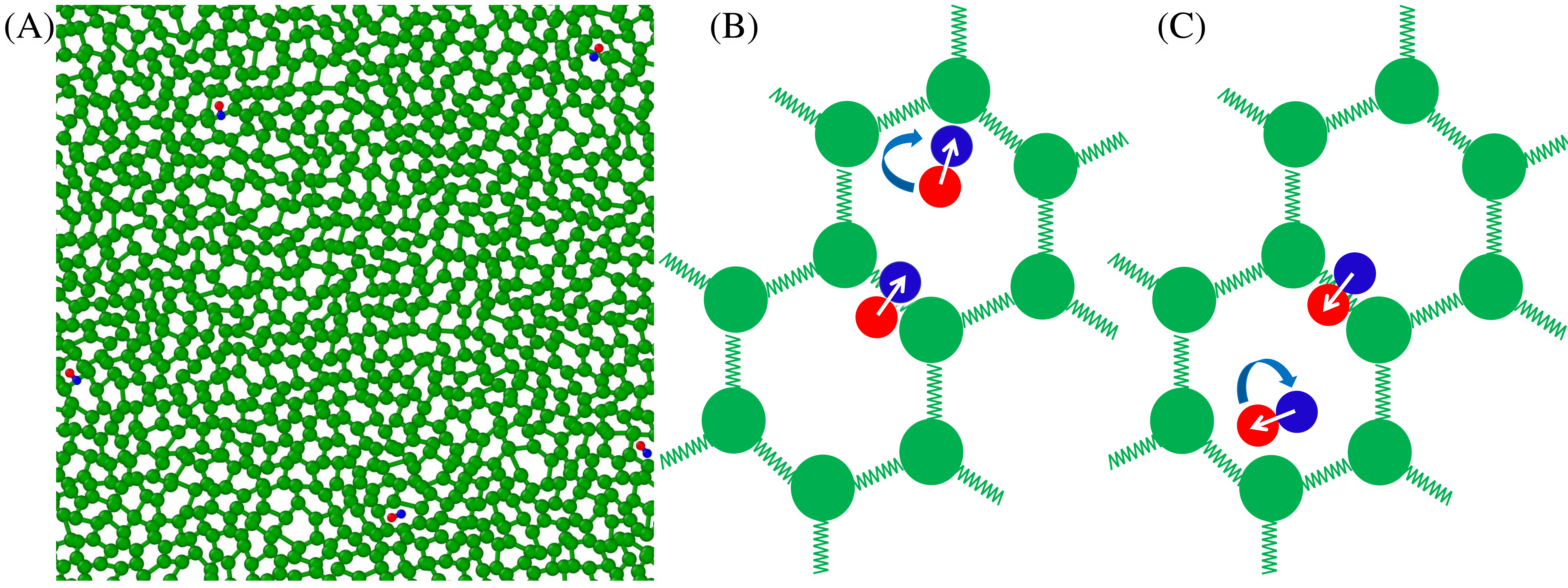}
    \caption{(A) A snapshot of the chemically asymmetric self-driven rigid dumbbells inside the polymer gel (green in color, springs are shown as solid lines). The snapshot is created with the Ovito package \cite{stukowski2009visualization}. A schematic representation of (B) trapping, rotation, and (C) mesh to mesh transition events of the self-driven rigid dumbbells in the polymer gel. The white arrows indicate the direction of self-propulsion.}
    \label{fig:2D_Gel}
\end{figure*}

\noindent We model the microswimmer as a rigid dumbbell by connecting two circular discs each of size 0.5 $\sigma$ $via$ harmonic spring potential.
\begin{equation}
V_{\text{harmonic}}\left(r_{ij} \right)= k_{\text{H}} \frac{(r_{ij} - r_{0})^2}{2}.
\label{eq:harm}
\end{equation}
\noindent Here, $k_{\text{H}} = 30$ is the spring constant and $r_{0}$ is the equilibrium distance between the two halves of the rigid dumbbell ($\delta l$=0.5$\sigma$), which is constant throughout the simulations and it ensures that the dimers are always translating and rotating as single rigid bodies. Next, the crowded environment is modeled by creating a  bead-spring polymer gel on a two-dimensional graphene lattice and each of these trivalent lattice sites are occupied by monomers of same diameter $\sigma$ (Fig.~\ref{fig:2D_Gel}A) and are connected to the neighboring monomers through the finitely extensible nonlinear elastic (FENE) spring potential defined as,
\begin{equation}
V_{\text{FENE}}\left(r_{ij} \right)=\begin{cases} -\frac{k r_{\text{max}}^2}{2} \log\left[1-\left( {\frac{r_{ij}}{r_{\text{max}}}}\right) ^2 \right],\hspace{5mm} \mbox{if } r_{ij} \leq r_{\text{max}}\\
=\infty, \hspace{41mm} \mbox{otherwise}
\end{cases}
\label{eq:FENE}
\end{equation}
where $r_{ij}$ is the distance between two neighboring monomers in the polymer gel with a maximum length $r_{\text{max}} = 2.5 \sigma$ and $k = 3$ is the force constant which accounts for the stiffness of the gel. The non-bonded gel monomer-monomer, two halves of the rigid dumbbells, and the rigid dumbbell with the monomers of the polymer gel interactions are set as purely repulsive and modeled by the Weeks–Chandler–Andersen (WCA) potential \cite{weeks1971role}: 
\begin{equation}
V_{\text{WCA}}(r_{ij})=\begin{cases}4\epsilon_{ij}\left[\left(\frac{\sigma_{ij}}{r_{ij}}\right)^{12}-\left(\frac{\sigma_{ij}}{r_{ij}}\right)^{6}\right]+\epsilon_{ij}, \hspace{1mm} \mbox{if } r_{ij} \leq (2)^{1/6}\sigma_{ij}\\
=0, \hspace{40mm} \mbox{otherwise}
\end{cases}
\label{eq:WCA}
\end{equation}
\noindent where $\sigma_{ij} = \frac{\sigma_i + \sigma_j}{2}$, with $\sigma_{i(j)}$ being the diameter of the interacting pairs. $\epsilon$ and $r_{ij}$ denote the interaction strength and the distance between the interacting particles. In our simulations, we model the rigid dumbbell as both the halves are attractive (sticky, blue colour in Fig.~\ref{fig:2D_Gel}) with the monomers of the polymer gel, which forms a chemically symmetric dumbbell. Further to make the dumbbell chemically asymmetric, we set the interactions in such a way that the two halves of the rigid dumbbells interacts differently with the monomers of the gel. For the asymmetric dumbbell one half interacts repulsively (Eq.~\ref{eq:WCA}) while the other half interacts attractively with the polymer gel (Fig.~\ref{fig:2D_Gel}). The attractive interaction is modeled $via$ a standard Lennard-Jones potential,
\begin{equation}
V_{\text{LJ}}(r_{ij})=\begin{cases}4\epsilon_{ij}\left[\left(\frac{\sigma_{ij}}{r_{ij}}\right)^{12}-\left(\frac{\sigma_{ij}}{r_{ij}}\right)^{6}\right], \hspace{5mm} \mbox{if } r_{ij} \leq r_\text{cut}\\
=0, \hspace{41mm} \mbox{otherwise}\\
\end{cases}
\label{eq:LJ}
\end{equation}
where $r_{ij}$ is the separation between one half of the rigid dumbbell and monomers of the gel, $\epsilon$ is the strength of the interaction (stickiness) with an interaction diameter $\sigma_{ij}$, and the Lennard-Jones cutoff length $r_{\textrm{cut}}$= $2.5$ $\sigma$. The Lennard-Jones parameters ($\sigma$ and $\epsilon$) and $m$ are the fundamental units of length, energy, and mass, respectively. Therefore, the unit of time is $ \tau = \sqrt{m\sigma^2/\epsilon}$.\\

\noindent The following Langevin equation is implemented to describe the dynamics of a particle with mass $m$ and position $r_{i}(t)$ at time $t$, interacting with all the other particles in the system: \\ 
\begin{equation}
m\frac{d^2 \textbf{r}_{i}(t)}{dt^2} = - \gamma \frac{d \textbf{r}_{i}}{dt} - \sum_{j} \bigtriangledown V(\textbf{r}_i-\textbf{r}_j) + {\bf f}_{i}(t) + {\bf{F}_{\text{a}} \bf \hat{n}} 
	\label{eq:langevineq}
\end{equation}
where $m$ is the mass of particles, $r_i$ is the position of $i^{th}$ particle at time $t$, $\gamma$ is the friction coefficient, and the total potential energy of the system can be written as $V(r) = V_{\text{FENE}} + V_{\text{harmonic}} + V_{\text{LJ}} + V_{\text{WCA}}$, where $V_{\text{FENE}}$ is spring potential for the polymer gel, $V_{\text{harmonic}}$ is spring potential connecting two halves of the rigid dumbbell, $V_{\text{LJ}}$ is the attractive potential and $V_{\text{WCA}}$ corresponds to excluded volume potential. We consider very high $\gamma$ and therefore for all practical purposes the dynamics is overdamped. Thermal fluctuations are captured by the Gaussian random force $f_i(t)$, satisfying the fluctuation-dissipation theorem.
\begin{equation}
	\left<f(t)\right>=0, \hspace{5mm}
	\left<f_{\alpha}(t^{\prime})f_{\beta}(t^{\prime\prime})\right>=4k_B T\gamma  \delta_{\alpha\beta}\delta(t^{\prime}-t^{\prime\prime})
	\label{eq:random-forcerouse}
	\end{equation}
\noindent where $k_{B}$ is the Boltzmann constant, T is the temperature and $ \delta$ represents the Dirac delta-function, $\alpha$ and $\beta$ represent the Cartesian components. We consider the thermal energy $k_B T=1$. The activity is modeled as a propulsive force $\text{F}_\text{a} \bf \hat {n}$, where $\text{F}_\text{a}$ represents the amplitude of active force with orientation specified by the unit vector $\bf \hat{n}$ connecting the centers of the two spherical discs which form the rigid dumbbell. The self-propulsion term $\text{F}_\text{a} \bf \hat {n}$ is zero for the passive polymer gel particles. In the case of asymmetric dumbbell we consider two directions of self-propulsion, either towards the sticky face or towards the non-sticky face of the dumbbell. Here, we express the activity in terms of a dimensionless quantity P\`{e}clet number, $\text{Pe}$  defined as $\text{Pe} = \frac{\text{F}_{\text{a}} \sigma}{k_B T}$. Therefore, $\text{Pe} = 0$ corresponds to the passive case. We do not include the hydrodynamic interactions in our simulations. \\

\noindent For each simulation, the system is initialized by randomly placing $5$ self-driven rigid dumbbells inside the polymer gel packed inside a square box of length $52 \sigma$ (Fig.~\ref{fig:2D_Gel}(A)). Periodic boundary conditions are set in all directions. All the simulations are performed using the Langevin thermostat, and the equation of motion (Eq.~\ref{eq:langevineq}) is integrated using the velocity Verlet algorithm in each time step. All the production simulations are carried out for $5 \times 10^8$ steps after relaxing the system long enough where the integration time step is considered to be $5 \times 10^{-4}$, and the positions of the particles are saved every $100^{th}$ step. This also ensures the equilibration of the polymer gel with an average mesh size $\sigma_\text{mesh} \approx 1.7 \sigma$ (Fig.~S1). We have performed $35$ independent simulations for each case. All the simulations were performed using LAMMPS \cite{plimpton1995fast}, a freely available open-source molecular dynamics package.

\section{Results and discussion}

\begin{figure*}
    \centering
    \includegraphics[width=0.9\linewidth]{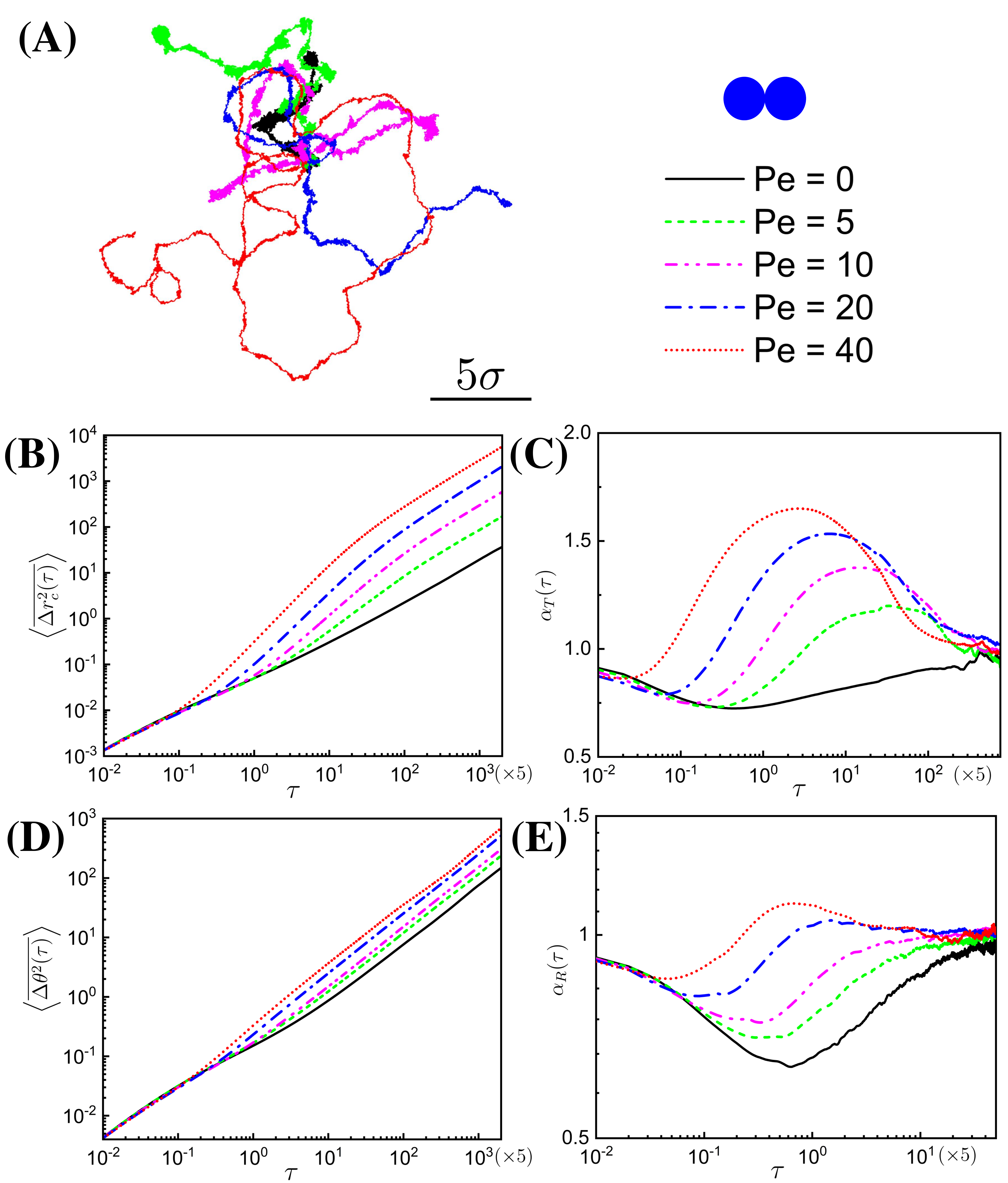}
  \caption{(A) The COM trajectories, log-log plot of (B) $\left<\overline{\Delta r_\text{c}^{2}(\tau)}\right>$ $vs$ $\tau$, log-linear plot of (C) $\alpha_{T}(\tau)$, log-log plot of (D) $\left<\overline{\Delta \theta^{2}(\tau)}\right>$ $vs$ $\tau$, and log-linear plot of (E) $\alpha_{R}(\tau)$ of the attractive ($\epsilon = 1.0$) self-driven symmetric dumbbell in 2D polymer gel for different Pe.}
    \label{fig:traj_TMSD_RMSD}
\end{figure*}

\noindent In order to study the dynamics of chemically symmetric and asymmetric self-driven rigid dumbbells inside the polymer gel, at the first step we compute the time-averaged translational and rotational mean square displacement, $\overline{\Delta_{i,\lbrace r_\text{c}, \theta \rbrace}^{2}(\tau)} = \frac{1}{T_\text{max}-\tau} \int_{0}^{T_\text{max}-\tau} \Delta_{i, \lbrace r_\text{c}, \theta \rbrace}^{2}(t, \tau)\, dt$, for all the initial time $t$ along each of the independent trajectories. Here, r$_\text{c}$ is the center of mass position, $\theta$ is the angle between x-axis and body axis of the dumbbell, and $T_\text{max}$ is the total length of the simulation. Next, the time-ensemble-average is obtained by $\left\langle{\overline{\Delta_{\lbrace r_\text{c}, \theta \rbrace}^{2}(\tau)}}\right\rangle  =  \frac{1}{\text{N}} \sum_{i=1}^{\text{N}}{\overline{\Delta_{i,\lbrace r_\text{c}, \theta \rbrace}^{2}(\tau)}}$, where $\text{N} = 35$ is the number of independent trajectories. Initially, we simulate passive and active rigid dumbbell in free space and analyze both the translational $\left(\left<\overline{\Delta r_\text{c}^{2}(\tau)}\right>\right)$ and rotational $\left(\left<\overline{\Delta \theta^{2}(\tau)}\right>\right)$ mean square displacements as a function of lag time $\tau$ to validate the parameters of our simulation (Fig.~S2). Note that for $\left<\overline{\Delta \theta^{2}(\tau)}\right>$, the cumulative angle $\theta$ is calculated from the simulation, therefore, the angle difference is no longer bounded by $2\pi$~\citep{klett2021non}. We fit the numerically calculated $\left<\overline{\Delta r_\text{c}^{2}(\tau)}\right>$ curves with the following analytical expression for an active Brownian particle \citep{volpe2014simulation}.
\begin{equation}
    \left<\overline{\Delta r_\text{c}^2(\tau)}\right> = \left[4D_{T} + v^{2} \tau_{R} \right] \tau + \frac{v^{2} \tau_{R}^{2}}{2}  \left[ e^{-\frac{2\tau}{\tau_{R}}} - 1 \right]
    \label{eq:analytical}
\end{equation}
Here $D_{T}$ is the thermal translational diffusion coefficient, $v$ is the self-propulsion velocity and $\tau_{R}$ is the persistence time. $\tau_{R} = \frac{1}{D_{R}}$, where $D_{R}$ is the thermal rotational diffusion coefficient (see supplementary for details).\\ 

\noindent Subsequently, to unravel the effect of crowding on the dynamics of the rigid dumbbells, we compute $\left<\overline{\Delta r_\text{c}^{2}(\tau)}\right>$ and $\left<\overline{\Delta \theta^{2}(\tau)}\right>$ for the self-driven dumbbells inside the polymer gel. We also calculate the translational and rotational time exponents defined as $\alpha_{T} (\tau) = \frac{\text{d log}\left<\overline{\Delta r_\text{c}^{2}(\tau)}\right>}{\text{d log}\tau}$, $\alpha_{R} (\tau) = \frac{\text{d log}\left<\overline{\Delta \theta^{2}(\tau)}\right>}{\text{d log}\tau}$ respectively. The translational and rotational motion of the self-driven dumbbells are strongly affected by the interactions with the monomers of the polymer gel. Initially, we consider a chemically symmetric dumbbell with both the halves sticky ($\epsilon = 1$) to the monomers of the polymer gel and study the effect of self-propulsion. We observe that the self-propulsion always leads to faster dynamics with increasing Pe as evident from the COM trajectory of the dumbbell shown in Fig.~\ref{fig:traj_TMSD_RMSD}A. $\left<\overline{\Delta r_\text{c}^{2}(\tau)}\right>$ shows a short time subdiffusion ($\alpha_{T} < 1$) (Fig.~\ref{fig:traj_TMSD_RMSD}C) due to the caging of the symmetric dumbbell inside the polymer mesh. This is caused because of the stickiness between the dumbbell and the polymer gel (Fig.~\ref{fig:traj_TMSD_RMSD}B) (Movie\_S1). At intermediate time the subdiffusive behavior becomes more pronounced for the passive (Pe = 0) case whereas in the case of self-driven symmetric dumbbell, $\left<\overline{\Delta r_\text{c}^{2}(\tau)}\right>$ grows faster with Pe in comparison to the passive dumbbell in polymer gel (Fig.~\ref{fig:traj_TMSD_RMSD}B). We find that the self-driven symmetric dumbbell also shows a subdiffusive behavior owing to the confined motion inside the polymer mesh for a very short time. As the time progresses, it starts to feel the effect of activity and performs a persistent motion, leading to superdiffusion ($\alpha_{T} > 1$) (Fig.~\ref{fig:traj_TMSD_RMSD}C) at the intermediate time (Fig.~\ref{fig:traj_TMSD_RMSD}B). This implies that the self-propulsion helps the symmetric dumbbell making a series of transitions between the meshes and explore a larger space inside the gel (Movie\_S2). At the longer time $\tau > \tau_{R}$, the direction of the self-propulsion is randomized due to the collisions and reorientation leading to a diffusive motion of the symmetric dumbbell, i.e. $\left<\overline{\Delta r_\text{c}^{2}(\tau)}\right>$ is linear in time with an enhanced diffusion coefficient. Apart from this, we also investigate the rotational dynamics of the symmetric dumbbell in the gel by analyzing $\left<\overline{\Delta \theta^{2}(\tau)}\right>$ (Fig.~\ref{fig:traj_TMSD_RMSD}D). Interestingly, we notice an activity-induced splitting of $\left<\overline{\Delta \theta^{2}(\tau)}\right>$ curves in Fig.~\ref{fig:traj_TMSD_RMSD}D for the symmetric dumbbell inside the polymer gel unlike the self-driven dumbbell in free space (Fig.~S2B). The self-driven dumbbell exhibits enhanced rotational dynamics with increasing Pe. At the intermediate time, $\left<\overline{\Delta \theta^{2}(\tau)}\right>$ exhibits a subdiffusive behavior ($\alpha_{R} < 1$) (Fig.~\ref{fig:traj_TMSD_RMSD}E), which is more prominent for the passive as well as for the smaller Pe values, and at higher Pe, the subdiffusion changes to superdiffusion ($\alpha_{R} > 1$) (Fig.~\ref{fig:traj_TMSD_RMSD}E). At longer time, $\left<\overline{\Delta \theta^{2}(\tau)}\right>$ displays an enhanced diffusive behaviour similar to Fig.~\ref{fig:traj_TMSD_RMSD}D. The self-driven dumbbell interacts more profoundly with the polymer gel and undergoes frequent collisions with the monomers of the gel. This results in an additional activity-induced torque responsible for the enhancement in rotation of the self-driven dumbbell compared to the passive dumbbell. Apart from this, if we increase the attraction strength ($\epsilon = 2$) of the symmetric dumbbell, $\left<\overline{\Delta r_\text{c}^{2}(\tau)}\right>$ and $\left<\overline{\Delta \theta^{2}(\tau)}\right>$ display qualitatively similar trends but the dynamics slows down because of the strong affinity of the dumbbell towards the polymer gel (Fig.~S3). Besides this, we also observe that the motion of self-driven rigid dumbbells become slower and strongly subdiffusive when the attraction ($\epsilon = 8$) between the dumbbell halves and gel is stronger than the energy scale associated with the self-propulsion (Pe = 5) (Fig.~S4). \\

\begin{figure*}
    \includegraphics[width=0.9\linewidth]{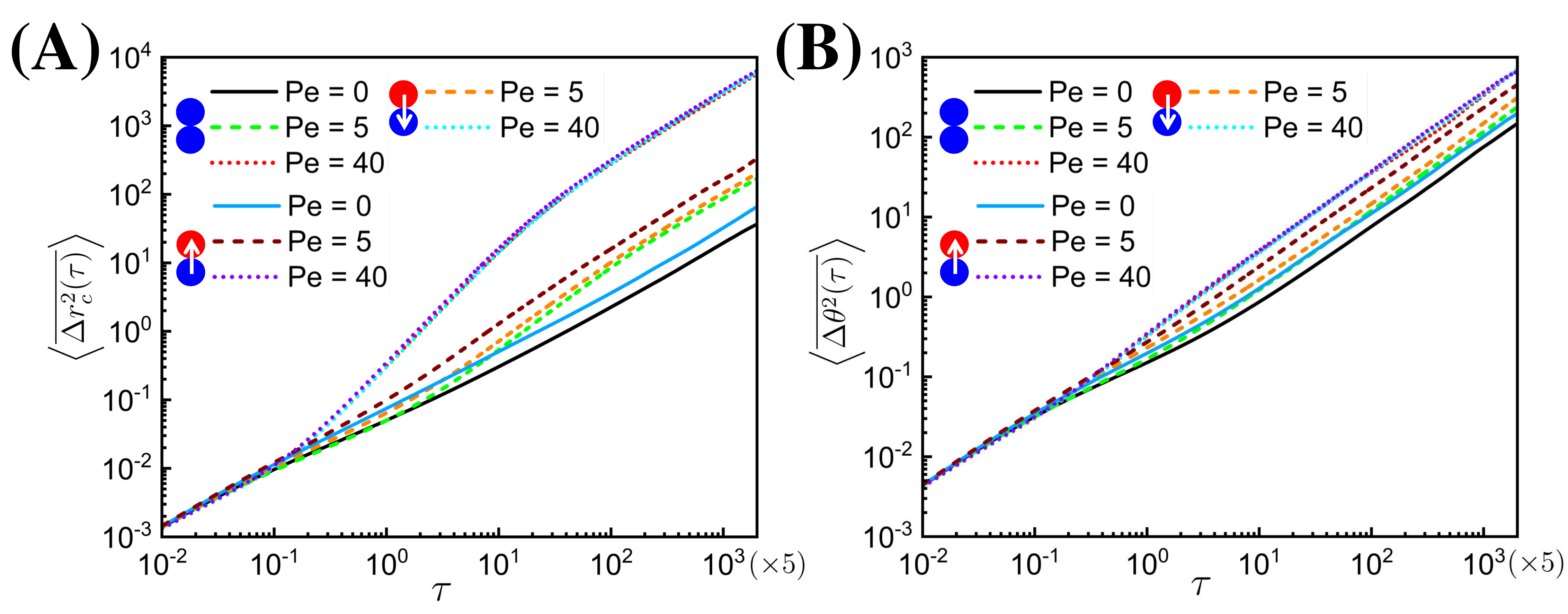}
    \caption{Log-log plot of (A) $\left<\overline{\Delta r_\text{c}^{2}(\tau)}\right>$ and (B) $\left<\overline{\Delta \theta^{2}(\tau)}\right>$ $vs$ the lag time $\tau$ of the passive and self-driven sticky ($\epsilon = 1.0$) symmetric dumbbell, asymmetric dumbbell with self-propulsion along the sticky half, and self-propulsion along the non-sticky half in 2D polymer gel for different Pe.}
    \label{fig:TMSD_RMSD_comp}
\end{figure*}

\noindent Further to understand the role of chemistry, as is mentioned in the introduction part, we make the rigid dumbbell chemically asymmetric by introducing the repulsive and sticky (attractive) interactions on the two different halves of the dumbbell. For the self-driven asymmetric dumbbell, the direction of self-propulsion plays a significant role in determining the dynamics inside the dense polymer gel. We consider two different directions of the self-propulsion along (a) the sticky half and (b) the non-sticky half of the asymmetric dumbbell. In Fig.~\ref{fig:TMSD_RMSD_comp}, we examine the effect of chemical asymmetry and the direction of self-propulsion by comparing it with the symmetric counterpart of the self-driven dumbbell for fixed values of self-propulsion (Pe = 0, 5, and 40) and a constant interaction strength $\epsilon = 1$. The $\left<\overline{\Delta r_\text{c}^{2}(\tau)}\right>$ and $\left<\overline{\Delta \theta^{2}(\tau)}\right>$ for the asymmetric dumbbell show enhancement in both translational and rotational dynamics with qualitatively similar trend compared to the symmetric dumbbell case for increasing Pe. For the asymmetric case, one half of the dumbbell interacts repulsively with the polymer gel which suppresses the possibility of getting stuck inside the polymer mesh unlike the symmetric dumbbell case, which gets trapped due to the higher effective sticky interaction experienced by both the two halves of the dumbbell possessing an attractive (sticky) interaction to the monomers of the gel. In addition to this, the direction of self-propulsion is also a key factor deciding the escape of the self-driven asymmetric dumbbell. We observe that the self-propulsion towards the sticky half promotes the dumbbell to interact more or stick with the polymer gel leading to transient trapping inside the polymer-mesh (Movie\_S3), while the self-propulsion towards the non-sticky half of the asymmetric dumbbell always facilitates the escape events from these traps (Movie\_S4). So the self-propulsion towards the non-sticky half of the asymmetric dumbbell exhibits enhanced translational and rotational dynamics in the polymer gel compared to the symmetric dumbbell and the self-propulsion directed along the sticky half of the asymmetric dumbbell\citep{theeyancheri2020translational}. At very high Pe, self-propulsion dominates and the chemical nature or the direction of self-propulsion is no longer have any effect on the dynamics of the dumbbell indicated by the overlapping of $\left<\overline{\Delta r_\text{c}^{2}(\tau)}\right>$ and $\left<\overline{\Delta \theta^{2}(\tau)}\right>$ curves in Fig.~\ref{fig:TMSD_RMSD_comp} for Pe = 40. The enhancement in rotational dynamics observed for the chemical asymmetric case is due to the extra torque arises from the additional force imbalance that comes from the simultaneous repulsive and attractive interactions experienced by the two half of the chemically asymmetric dumbbell compared to the symmetric dumbbell.  \\

\begin{figure*}
    \centering
    \includegraphics[width=0.78\linewidth]{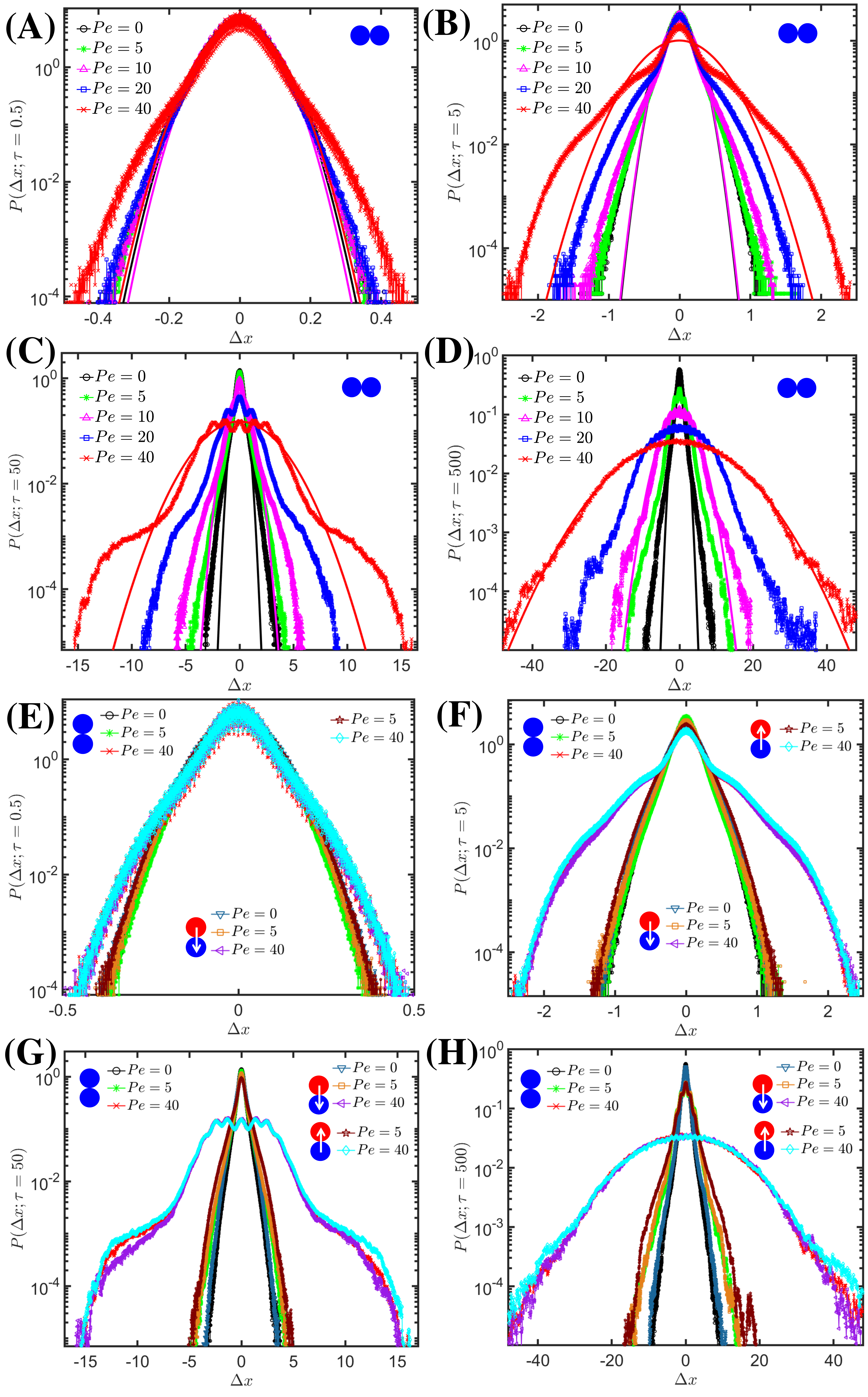}
    \caption{Plots of $\text{P}(\Delta x; \tau)$ for the symmetric sticky ($\epsilon = 1.0$) dumbbell as a function of Pe at different lag-times $\tau$ (0.5, 5, 50 and 500) (A-D). Comparison of $P(\Delta x; \tau)$ for asymmetric dumbbell with self-propulsion along the sticky ($\epsilon = 1.0$) half (arrow towards blue half), self-propulsion along the non-sticky half (arrow towards red half), and symmetric dumbbell in the polymer gel for Pe = 0, 5, and 40 at different lag-times $\tau$ (0.5, 5, 50 and 500) (E-H). The Solid lines represent the Gaussian fittings.}
    \label{fig:PDF_sym_asym}
\end{figure*}

\noindent In order to achieve deeper insights of the dynamics of the self-driven dumbbell in the polymer gel, we analyze the self part of the van Hove correlation function (probability distribution function) $P(\Delta x; \tau) \equiv \left\langle \delta(\Delta x - (x(t+\tau)-x(t)))\right\rangle$ of the dumbbell displacement in one dimension (along x-direction, see Fig.~S5 for along y-direction), where $x(t+\tau)$ and $x(t)$ are the center of mass positions of the dumbbell along x-direction at time $(t+\tau)$ and t respectively. Fig.~\ref{fig:PDF_sym_asym} depicts the $P(\Delta x; \tau)$ with the corresponding Gaussian distribution fittings for the Brownian motion, $P (\Delta x; \tau)=\frac{1}{\sqrt{2\pi\left<\Delta x^2\right>}}\exp\left({-\frac{\Delta x^2}{2\left<\Delta x^2\right>}}\right)$ for different lag-time $\tau$. Firstly, we focus on the symmetric dumbbell which interacts attractively with the polymer gel, and study the effect of self-propulsion on $P(\Delta x; \tau)$ by varying Pe (Fig.~\ref{fig:PDF_sym_asym}(A, B, C, and D)). At very short time for smaller values of Pe, all the curves are merged, and the larger Pe shows a comparatively broader distribution for $P(\Delta x; \tau)$ (Fig.~\ref{fig:PDF_sym_asym}A). As the time progresses, the dumbbells cover more space inside the gel, and $P(\Delta x; \tau)$ becomes broader with increasing Pe at the intermediate time (Fig.~\ref{fig:PDF_sym_asym}(B, C)). More surprisingly, there exists multiple central peaks in $P(\Delta x; \tau)$ at the intermediate time with increasing Pe as shown in Fig.~\ref{fig:PDF_sym_asym}C. This behavior is a consequence of the adjacent mesh to mesh motion of the self-driven dumbbell. Typically, an active (self-propelled) particle in confinement tends to move towards the boundary of the confinement, and escapes if given a chance. Therefore, the occurrence of these multiple central peaks in $P(\Delta x; \tau)$ is a characteristic of the such mesh to mesh motion of the self-driven dumbbell. Similar multiple central peaks in the van Hove function are reported earlier in theoretical studies of tracer particle dynamics in an elastic active gel~\citep{ben2015modeling, razin2019signatures}. The self-propulsion further facilitates the mesh to mesh motion of the symmetric dumbbell, and the number of central peaks increases with Pe, which manifests that the self-driven dumbbell explores larger number of meshes in the gel. In the long-time limit, the distribution becomes broader owing to the diffusive motion as shown in Fig.~\ref{fig:PDF_sym_asym}D. Moreover, we find that the probability density profiles can be fitted with Gaussian distributions (solid lines in Fig.~\ref{fig:PDF_sym_asym}(A-D)) for passive dumbbell at the short time scale, whereas $P(\Delta x; \tau)$ starts deviating from the Gaussianity with increasing Pe owing to the self-driven motion of the dumbbell. The deviation from the Gaussianity is more pronounced at the intermediate time for higher Pe. In the long time when the dumbbell reaches the diffusive regime, $P(\Delta x; \tau)$ shows the Gaussianity (Fig.~\ref{fig:PDF_sym_asym}(A-D)). Apart from this, if we make the symmetric dumbbell more sticky ($\epsilon = 2.0$), it gets harder to move from one mesh to other mesh as the dumbbell experiences a strong affinity towards the monomers of the gel. This leads to narrower $P(\Delta x; \tau)$ with sharp peaks compared to the less sticky ($\epsilon = 1.0$) symmetric dumbbell. However, the qualitative behavior remains the same (Fig.~S6). We also notice that $\text{P}(\Delta x; \tau)$ exhibits a very narrow distribution in the case where the stickiness ($\epsilon = 8$) is stronger than the energy scale associated with the self-propulsion force (Pe = 5) (Fig.~S7). In this case, the dumbbell possess a very strong affinity towards the gel and thus it stays inside the same mesh for a longer time. Next, we examine whether the effects of chemical asymmetry and direction of self-propulsion modify the probability density profile of the self-driven dumbbell by computing $P(\Delta x; \tau)$ for the chemically asymmetric dumbbell. We notice a qualitatively similar behavior in $P(\Delta x; \tau)$ for the asymmetric dumbbell as in case of the symmteric one (Fig.~\ref{fig:PDF_sym_asym}(E-H)). For the passive case and smaller values of Pe, there exists a clear difference in $P(\Delta x; \tau)$ as it displays a broader distribution for the asymmetric case with self-propulsion along the repulsive half of the dumbbell in comparison to the symmetric and self-propulsion along sticky half of the asymmetric dumbbell (Fig.~\ref{fig:PDF_sym_asym}(F, G)). This difference is more prominent at the intermediate time. The self-propulsion along the sticky face promotes trapping, in contrast, the self-propulsion along the non-sticky half facilitates the escape of asymmetric dumbbell from the transient trapping while moving through the polymer gel. This results a broader probability density profile for asymmetric dumbbell with the self-propulsion directed towards the non-sticky half (Fig.~\ref{fig:PDF_sym_asym}F). At higher Pe, the self-propulsion dominates over the effect of chemical asymmetry and the self-propulsion direction. Therefore, the distribution curves merge and fall on top of each other.

\section{Conclusions}

\noindent In this work, we present an extensive computational study of the translational and rotational dynamics of passive as well as the chemically symmetric and asymmetric self-driven rigid dumbbells inside a polymer gel in two-dimension. Our analyses manifest that the passive symmetric dumbbell shows a more pronounced subdiffusion at the intermediate time because of the confinements inside the the polymer mesh. On the other hand, the self-driven symmetric dumbbell exhibits enhanced translational and rotational dynamics with a superdiffusion at the intermediate time and an enhanced diffusion at long-time as a function of increasing Pe. This implies that self-propulsion always enhances the translational and rotational dynamics by effectively facilitating the escape of symmetric dumbbells from the transient trapping due to polymer mesh. Moreover, the additional torque coming from the activity-induced interactions results in enhanced rotational dynamics. To achieve a deeper understanding of the chemical interactions controlling the dynamics in crowded media, we examine the transport of an asymmetric dumbbell in the polymer gel. Our results display that the direction of self-propulsion plays a significant role in the translational and rotational dynamics of the asymmetric dumbbell. In addition to that, the chemical asymmetry leads to faster dynamics compared to the symmetric case. The direction of self-propulsion is crucial in the context of asymmetric dumbbells, the self-propulsion along the sticky half promotes the trapping while the propulsion directed along the non-sticky half facilitates the escape of the self-driven dumbbell from the transient trapping caused by the polymer-mesh. Also, an enhanced rotational motion results from the extra torque caused due to the force imbalance created by the different chemical interactions by the two halves of the dumbbell with the polymer gel for the asymmetric case. These findings are further supported by the van Hove correlation functions which display increasingly broader distribution with increasing Pe, indicating the larger displacement of the self-driven dumbbell as a function of increasing self-propulsion. Both the symmetric and asymmetric self-driven dumbbells manifest qualitatively similar behavior but the asymmetric case where the self-propulsion acts along non-sticky half gives a broader distribution than the asymmetric case with self-propulsion acts along sticky half and the symmetric dumbbell at the intermediate time for the smaller values of Pe. The dynamics of the self-driven dumbbells display deviations from the Gaussianity for a higher range of activities at the intermediate time, whereas the dynamics remains as Gaussian at long-time for higher Pe. Our present study depicts how the dynamics and transport of self-driven dumbbells in polymer gel depend on the direction of self-propulsion and the symmetric or asymmetric chemical interaction with the environments. \\

\noindent In brief, the present work portrays the dynamics of the self-driven dumbbells in polymer gel in two-dimension. It is already reported that crowding significantly affects biomolecular transport \citep{lieleg2012mucin, hofling2013anomalous}. The mesh-like environments exist in various systems: the intercellular space contains the mesh-like extracellular matrix in cellular tissue compartments~\cite{mclane2013spatial, ghosh2016interactions, petrova2017conformations}. In novel clinical diagnosis tools, pathogens diffuse in  hydrogels~\citep{shin2014sensing}. So it is biologically important to understand the diffusion through such a complex and crowded gel-like medium. Apart from this, to understand the active transport of molecules through these dense space, researchers have designed different artificial swimmers like the Janus rods \citep{patra2013intelligent}, chiral particles \citep{ghosh2009controlled}, and vesicles\citep{joseph2017chemotactic} which mimic the real biological transporters. These artificial microswimmers are used in nanomedicine for site-specific delivery of cargo and also the self-powered agents are considered as next-generation drug delivery vehicles~\citep{patra2013intelligent}. So they are subjected to a range of topological constraints and interactions, while performing the assigned tasks. It is highly demanding to have a thorough understanding and tuning of highly selective transport by these artificial agents through crowded regions such as a biological cell. We hope that the main findings reported in our work stay valid in providing insights on active diffusion of microswimmers in crowded media. We believe that our investigation will help scientists and engineers to design efficient, powerful, and better performing artificial self-driven transport machines in crowded environments. We also hope that our proposed model will motivate experimentalists to design and perform experiments and testify our predictions in near future by synthesizing chemically asymmetric self-powered dumbbell shaped colloids.

\section*{Acknowledgments}

\noindent P. K. and L. T. thank UGC for the fellowships. R. C. acknowledges SERB for funding (Project No. MTR/2020/000230 under MATRICS scheme) and IRCC-IIT Bombay (Project No. RD/0518-IRCCAW0-001) for funding. The authors thank Dr. Subhasish Chaki for helpful discussions and critically reading the manuscript. P. K. would like to thank Dr. Koushik Goswami for scientific discussions. The authors acknowledge Sanaa Sharma for reading the manuscript. P.K. thanks Kailash Sahu for the help. \\

\noindent \textbf{DATA AVAILABILITY} \\

\noindent The data that supports the findings of this study are available within the article [and its supplementary material]. \\



\renewcommand{\thefigure}{S\arabic{figure}}
\setcounter{figure}{0}
\setcounter{equation}{0}
\renewcommand{\theequation}{\Roman{equation}}
\appendix
\noindent \textbf{SUPPLEMENTARY MATERIAL} \\

\begin{figure*}
    \centering
    \includegraphics[width=0.95\linewidth]{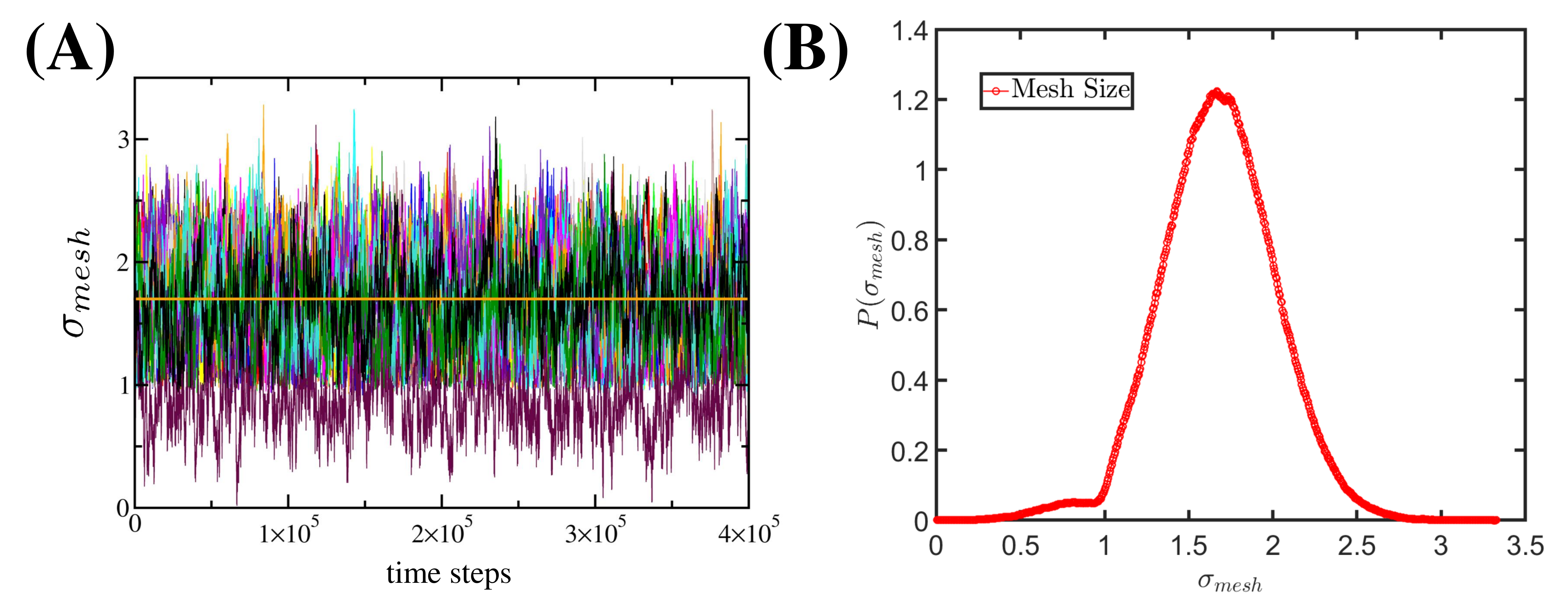}
    \caption{Plots of (A) the mesh size fluctuation ($\sigma_{mesh}$) \textit{vs} time steps with an average mesh size $\sim 1.7$ (yellow solid line) and (B) mesh size distribution P($\sigma_{mesh}$) of the polymer gel.}
    \label{fig:S1_Mesh_size_fluc_dist}
\end{figure*}

\noindent To characterize the polymer gel, we measure the average mesh size of the polymer gel by considering the diagonal distances between the monomers of the hexagons which is the basic repeating unit of the polymer gel as shown in Fig.~\ref{fig:S1_Mesh_size_fluc_dist}.

\begin{figure*}
    \centering
    \includegraphics[width=0.95\linewidth]{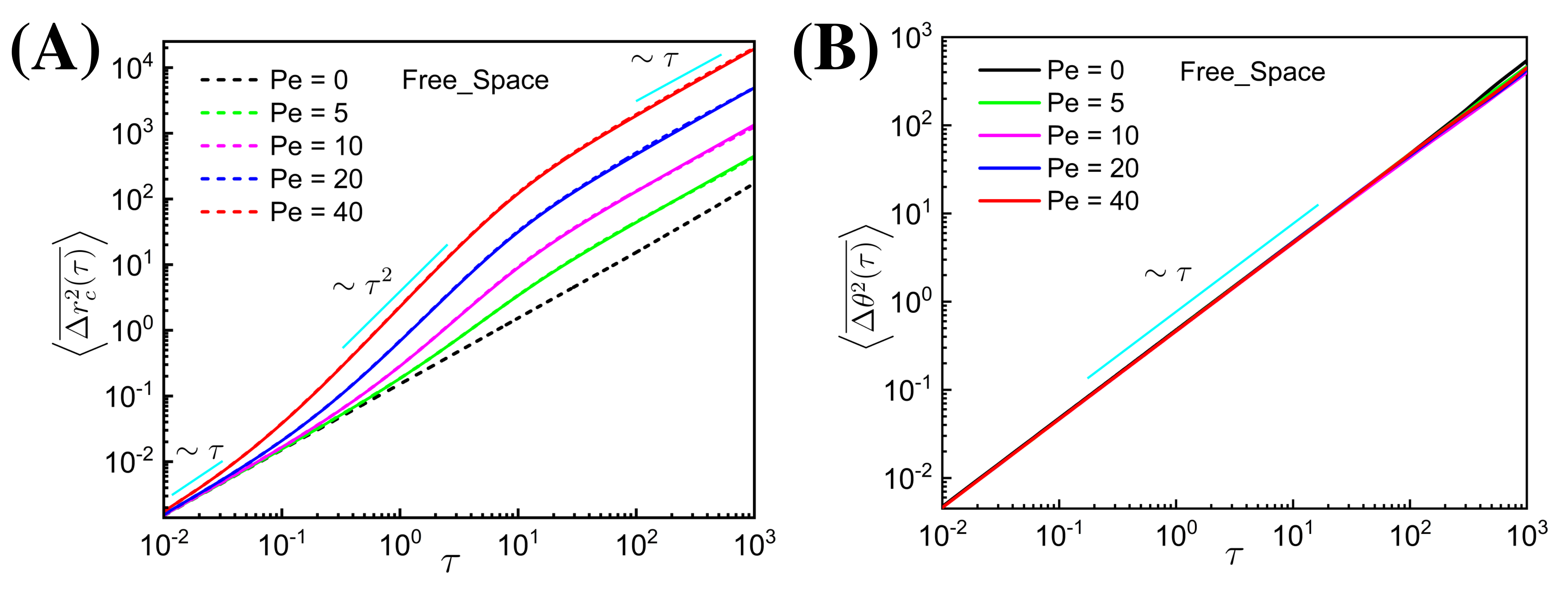}
    \caption{log-log plot of (A) $\left<\overline{\Delta r_\text{c}^{2}(\tau)}\right>$ fitted with Eq.~(\ref{eq:Analytical}) (solid lines) and (B) $ \left<\overline{\Delta \theta^{2}(\tau)}\right>$ for the self-propelled rigid dumbbell in free space at different Pe.}
    \label{fig:S2_TMSD_RMSD_Free}
\end{figure*}

\noindent The method and parameter validation is carried out for a rigid dumbbell in free space. The translational $\left(\left<\overline{\Delta r_\text{c}^{2}(\tau)}\right>\right)$ and rotational $\left(\left<\overline{\Delta r_\theta^{2}(\tau)}\right>\right)$ mean square displacement are calculated. From the plot (Fig.~\ref{fig:S2_TMSD_RMSD_Free}) for Pe = 0, we have computed the thermal translational diffusion coefficient, $D_{T} = 3.79 \times 10^{-4}$ and the rotational diffusion coefficient, $D_R = 2.40 \times 10^{-3}$. Thus, the persistence time $\tau_{R} = \frac{1}{D_{R}} = 4.16 \times 10^{2}$. The self-propulsion velocity is $v = \frac{F_a}{\gamma}$. Using the values of $D_T$ and $\tau_R$, $\left<\overline{\Delta r_\text{c}^{2}(\tau)}\right>$ is fitted with the analytical expression for an active Brownian particle: 
\begin{equation}
    \left<\overline{\Delta r_\text{c}^2(\tau)}\right> = \left[4D_{T} + v^{2} \tau_{R} \right] \tau + \frac{v^{2} \tau_{R}^{2}}{2}  \left[ e^{-\frac{2\tau}{\tau_{R}}} - 1 \right]
    \label{eq:Analytical}
\end{equation}

\noindent For the passive rigid dumbbell (Pe = 0), $\left<\overline{\Delta r_\text{c}^{2}(\tau)}\right>$ is always diffusive $\left(\left<\overline{\Delta r_\text{c}^{2}(\tau)}\right> \sim \tau\right)$ in whole range of time with the diffusion coefficient $D_{T}$. In case of self-driven rigid dumbbell, $\left<\overline{\Delta r_\text{c}^{2}(\tau)}\right>$ exhibits three distinct regions: diffusive at very short time ($\tau << \tau_{R}$) with the analytical expression $\left<\Delta r_\text{c}^2(\tau)\right> = 4D_{T} \tau $, a superdiffusive region at intermediate time ($\tau \simeq \tau_{R}$) which scales as $\left<\overline{\Delta r_\text{c}^{2}(\tau)}\right>$ $\sim \tau^2$, and the expression becomes $\left<\Delta r_\text{c}^2(\tau)\right> = 4D_{T} \tau + 2 v^{2} \tau^2$, followed by enhanced diffusive region at longer time, \textit{i.e.} $\tau >> \tau_{R}$ and the expression becomes $\left<\Delta r_\text{c}^2(\tau)\right> = (4D_{T} + 2 v^{2} \tau_{R})\tau$. $\left<\overline{\Delta r_\text{c}^{2}(\tau)}\right>$ grows faster with Pe in comparison to the passive dumbbell (shown in Fig.~\ref{fig:S2_TMSD_RMSD_Free}A). On the other hand, $\left<\overline{\Delta \theta^{2}(\tau)}\right>$ of rigid dumbbell remains independent of self-propulsion in free space \textit{i.e.} motion is diffusive with constant rotational diffusion coefficient $D_{R} = 2.40 \times 10^{-3}$, shown in Fig.~\ref{fig:S2_TMSD_RMSD_Free}B, which depicts that the persistence time $\tau_{R} = 4.16 \times 10^{2}$ is same for all the curves with different Pe in free space. It is due to the fact that in free space the rotation is governed only by thermal fluctuations. \\

\begin{figure*}
    \centering
    \includegraphics[width=0.95\linewidth]{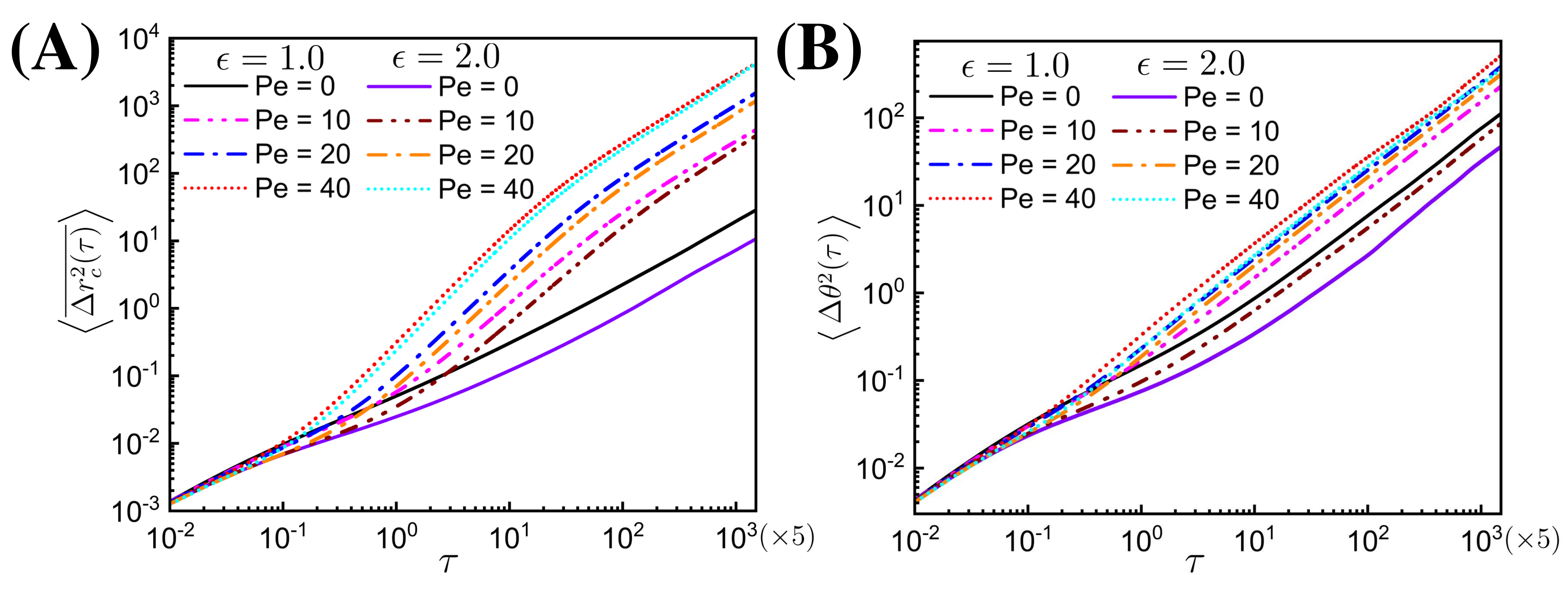}
    \caption{log-log plots of (A) $\left<\overline{\Delta r_\text{c}^{2}(\tau)}\right>$ $vs$ $\tau$, and (B) $\left<\overline{\Delta \theta^{2}(\tau)}\right>$ $vs$ $\tau$ for the attractive ($\epsilon$ = 1.0 and 2.0) self-driven symmetric dumbbell in polymer gel at different Pe.}
    \label{fig:S3_TMSD_RMSD_E1E2}
\end{figure*}

\begin{figure*}
    \centering
    \includegraphics[width=0.95\linewidth]{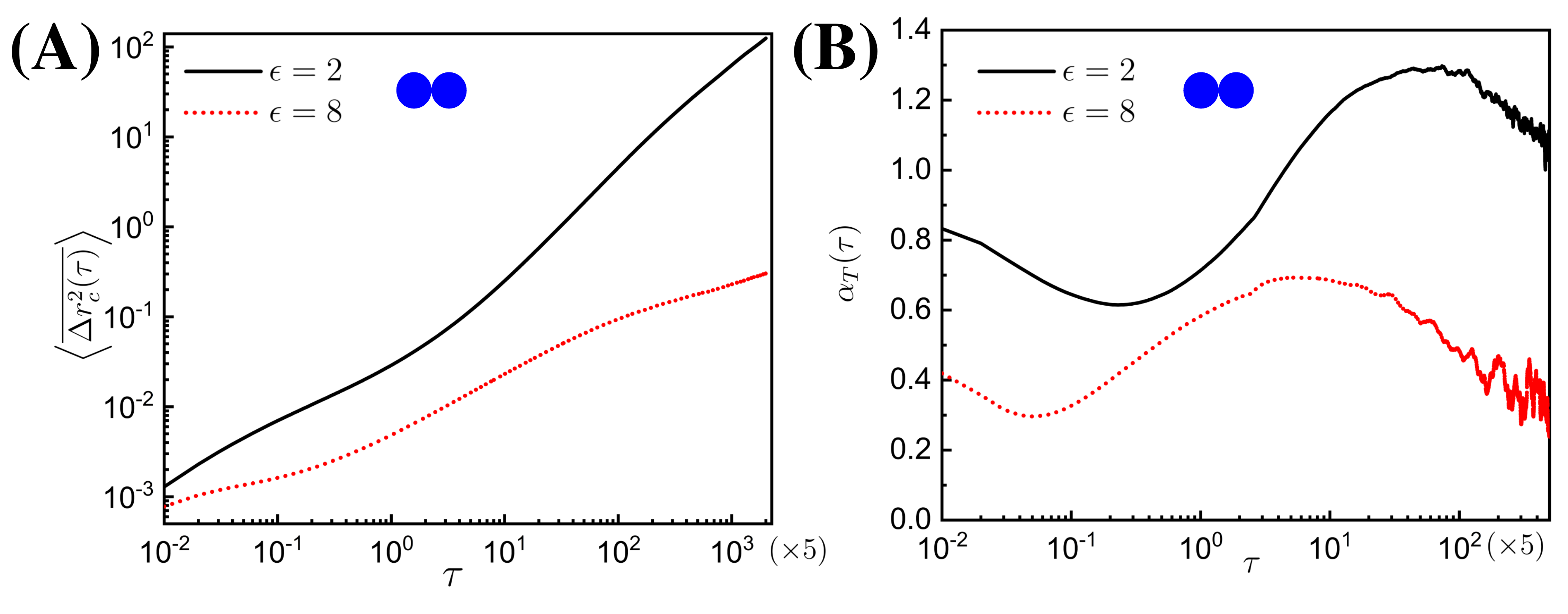}
    \caption{log-log plot of (A) $\left<\overline{\Delta r_\text{c}^{2}(\tau)}\right>$ $vs$ $\tau$ and log-linear plot of (B) $\alpha_{T}(\tau)$ $vs$ $\tau$ for the attractive ($\epsilon$ = 2.0 and 8.0) self-driven symmetric dumbbell in polymer gel at Pe = 5.}
    \label{fig:S4_MSD_Exp_Epsilon}
\end{figure*}

\begin{figure*}
    \centering
    \includegraphics[width=0.75\linewidth]{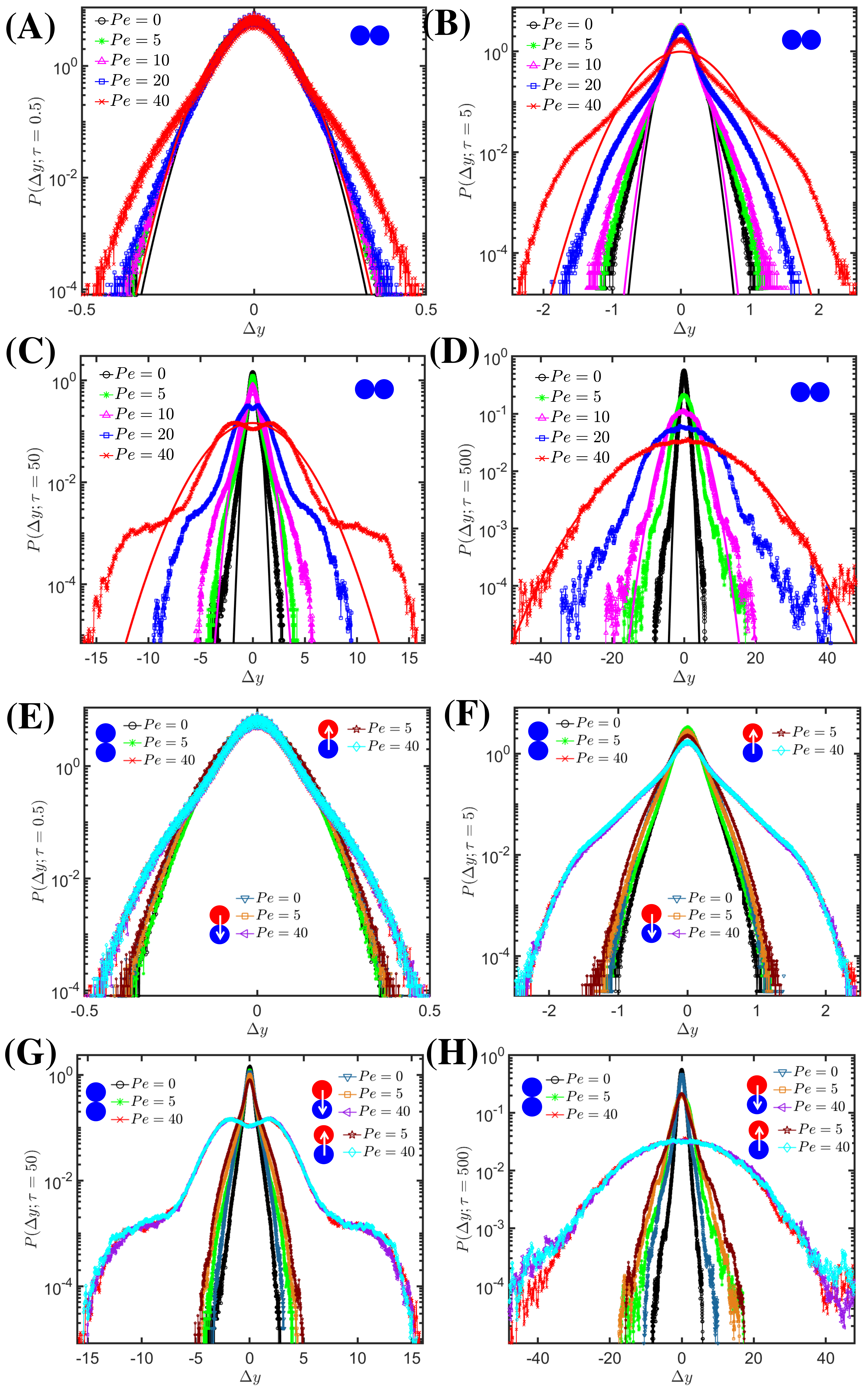}
    \caption{Plots of $\text{P}(\Delta y; \tau)$ for the symmetric sticky ($\epsilon = 1.0$) dumbbell as a function of Pe at different lag-times $\tau$ (0.5, 5, 50 and 500) (A-D). Comparison of $P(\Delta y; \tau)$ for asymmetric dumbbell with self-propulsion along the sticky ($\epsilon = 1.0$) half (arrow towards blue half), self-propulsion along the non-sticky half (arrow towards red half), and symmetric dumbbell in the polymer gel for Pe = 0, 5, and 40 at different lag-times $\tau$ (0.5, 5, 50 and 500) (E-H). The Solid lines represent the Gaussian fittings.}
    \label{fig:S5_PDF_dY}
\end{figure*}

\begin{figure*}
    \centering
    \includegraphics[width=0.95\linewidth]{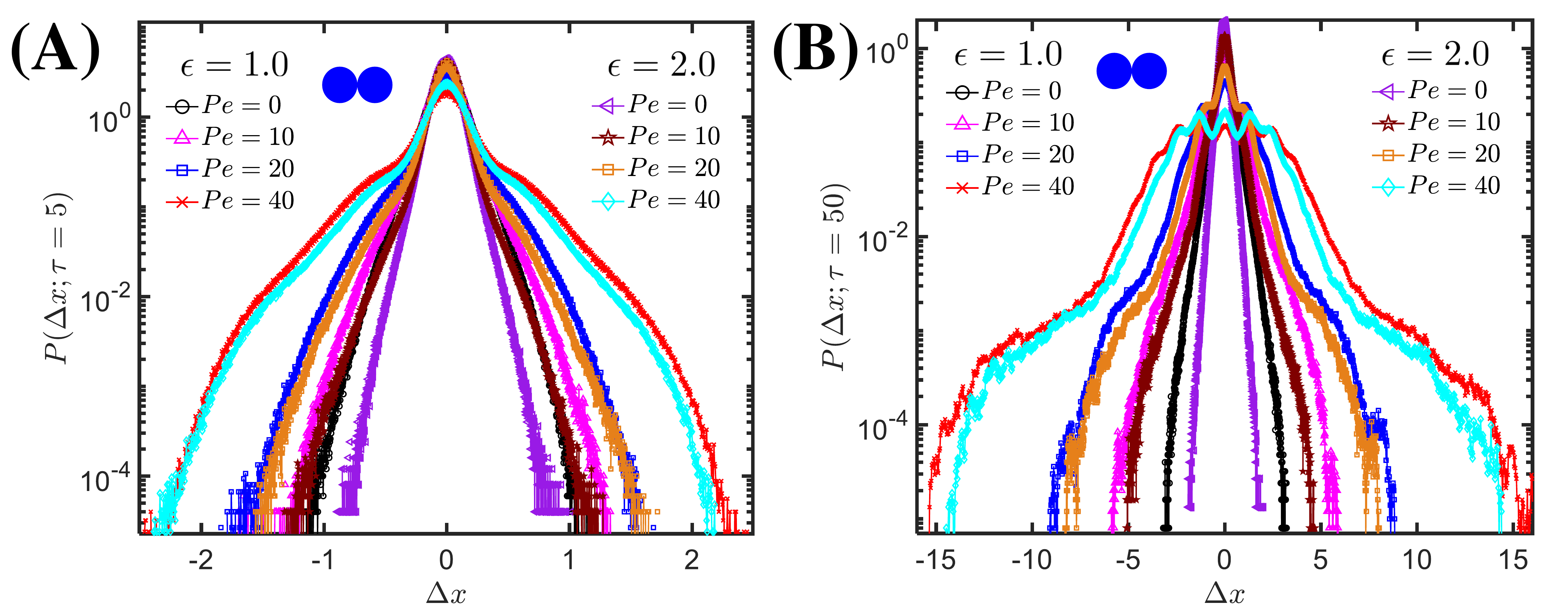}
    \caption{$P(\Delta x; \tau)$ for the attractive symmetric dumbbell for $\epsilon = 1.0$ and $\epsilon = 2.0$ at lag-time (A) $\tau = 5$, and $\tau = 50$ for different Pe.}
    \label{fig:S6_PDF_E1E2}
\end{figure*}

\begin{figure*}
    \centering
    \includegraphics[width=0.95\linewidth]{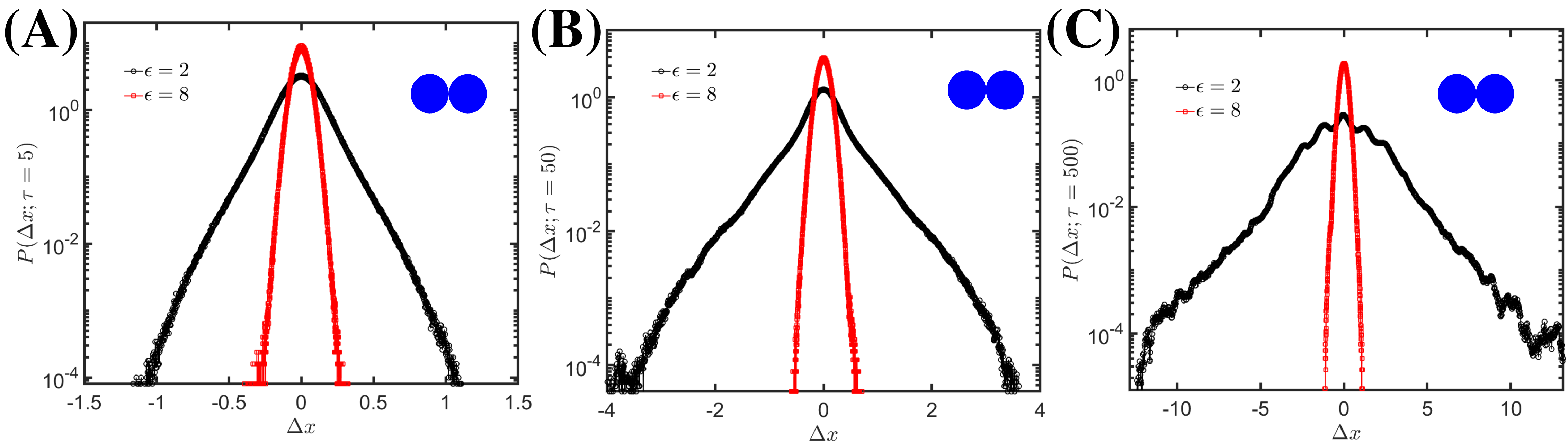}
    \caption{$P(\Delta x; \tau)$ for the attractive symmetric dumbbell for $\epsilon = 2.0$ and $\epsilon = 8.0$ at lag-time (A) $\tau = 0.5, 5$, and $\tau = 50$ for Pe = 5.}
    \label{fig:S7_PDF_E2E8F5}
\end{figure*} 

\noindent \textbf{Movie description} \\

\noindent Movie\_S1 \\
\noindent Molecular dynamics simulation of the passive (Pe = 0) chemically symmetric rigid dumbbells with both the halves are sticky to the monomers of the polymer gel (green in color). It is clear from the movie that the chemically symmetric rigid dumbbells are transiently trapped inside the polymer mesh. \\

\noindent Movie\_S2 \\
\noindent Molecular dynamics simulation of the self-driven (Pe = 20) chemically symmetric rigid dumbbells with both the halves are sticky to the monomers of the polymer gel. One can see that the self-propulsion helps the dumbbell escape from the confined polymer mesh and cover a larger area inside the gel. \\

\noindent Movie\_S3 \\
\noindent Molecular dynamics simulation of the self-driven (Pe = 20) chemically asymmetric rigid dumbbells with one sticky half (blue) and the other non-sticky (red) to the monomers of the polymer gel. Here, the direction of the self-propulsion is along the sticky (blue) half, which promotes the hindered motion of dumbbells inside the gel as it moves along the sticky half. \\

\noindent Movie\_S4 \\
\noindent Molecular dynamics simulation of the self-driven (Pe = 20) chemically asymmetric rigid dumbbells with one half being sticky (blue) while the other half being non-sticky (red) to the monomers of the polymer gel. Here, the direction of the self-propulsion is along the non-sticky half (red), which facilitates the escape of the dumbbells as it moves along the non-sticky half inside the gel.

\clearpage

\providecommand{\noopsort}[1]{}\providecommand{\singleletter}[1]{#1}%

\end{document}